\documentclass[a4paper,11pt]{article}
\usepackage{pos}

\title{Analysis Description Languages for the LHC}

\author*[a]{Sezen Sekmen}
\author[b]{Philippe Gras}
\author[c]{Lindsey Gray}
\author[d]{Benjamin Krikler}
\author[e]{Jim Pivarski}
\author[f]{Harrison B. Prosper}
\author[g]{Andrea Rizzi}
\author[h]{Gokhan Unel}
\author[i]{Gordon Watts}

\affiliation[a]{Center for High Energy Physics, Kyungpook National University, Daegu, South Korea}
\emailAdd{ssekmen@cern.ch}

\affiliation[b]{CEA Institut de Recherche sur les lois Fondamentales de l’Univers, Université Paris-
Saclay, Gif-sur-Yvette, France}

\affiliation[c]{Fermi National Accelerator Laboratory
Batavia, IL 60510, USA}

\affiliation[d]{H.H. Wills Physics Laboratory, University of Bristol, Bristol, United Kingdom}

\affiliation[e]{Department of Physics, Princeton University, Princeton, NJ, USA}

\affiliation[f]{Department of Physics, Florida State University, Tallahassee, FL, USA}

\affiliation[g]{INFN and University of Pisa, Italy}

\affiliation[h]{Physics and Astronomy Department, University of California at Irvine, Irvine, CA, USA}

\affiliation[i]{Department of Physics, University of Washington, Seattle, WA, USA}

\abstract{An analysis description language is a domain specific language capable of describing the contents of an LHC analysis in a standard and unambiguous way, independent of any computing framework. It is designed for use by anyone with an interest in, and knowledge of, LHC physics, i.e., experimentalists, phenomenologists and other enthusiasts.  Adopting analysis description languages would bring numerous benefits for the LHC experimental and phenomenological communities ranging from analysis preservation beyond the lifetimes of experiments or analysis software to facilitating the abstraction, design, visualization, validation, combination, reproduction, interpretation and overall communication of the analysis contents.  Here, we introduce the analysis description language concept and summarize the current efforts ongoing to develop such languages and tools to use them in LHC analyses.}

\FullConference{%
  The Eighth Annual Conference on Large Hadron Collider Physics-LHCP2020 \\
  25-30 May, 2020\\
  online}

\begin{document}
\maketitle



The unprecedented amount of data collected by the LHC likely carries hints for physics beyond the standard model. LHC physicists are thus designing an ever growing number and diversity of elaborate analyses to discover those hints.  These analyses focus on different final states involving different types of objects with varying properties and different kinematic variables upon which varying selections are applied. Exploring data with hundreds of different analyses increases our chances of discovery, but also brings many complexities and challenges.  These include working with many alternative definitions for objects, event variables with ambiguous definitions, and cases with hundreds of selection regions.  It becomes difficult to get a complete view of which signatures are covered, which ones are not, which analyses have disjoint subsets and which ones have overlapping subsets.  It is therefore critical to find ways of expressing and presenting analysis information in a most clear and organized manner.  In this report, we present the analysis description language approach, which is recently being studied for these purposes.

An analysis description language for the LHC is a special language capable of describing an LHC analysis in a standard and unambiguous way. It is customized to express analysis-specific concepts, and is designed for use by anyone with an interest in and knowledge of LHC physics, including experimentalists, phenomenologists or other enthusiasts. The principles of an analysis description language were first thoroughly discussed within the HEP community during the Les Houches PhysTeV 2015 workshop~\cite{Brooijmans:2016vro}.  It was suggested that the language should be complete in content, demonstrably correct, easily learnable and sustainable.  Moreover, it would be desirable to have a language that is easy to read and self-contained.  LHC analyses are typically performed using a large diversity of analysis frameworks custom-designed by each analysis group.  These frameworks are structurally complex, and the physics information is scattered in multiple components, which leads to a steep learning curve and difficulty of preservation.  Therefore it is also crucial to explore options that are framework-independent.  The design could target a domain specific language (DSL), or an embedded DSL (which is a DSL based on the syntax of a general purpose language (GPL)).  A practical option would be to build a declarative language, which can express the logic of a computation without describing its control flow. 

By construction, an analysis description language is not designed to be general purpose, therefore, getting the right scope is key. Languages can be developed to target different analysis subsystems such as event processing, histogramming and visualization, fitting and statistical inference, or workflow management.  Current studies have mainly focused on event processing and histogramming.  For completeness, the core of event processing would include definitions of simple and composite objects, event variables and event selections.

Working with analysis description languages could benefit experimentalists, phenomenologists and even the broader public interested in physics.  It would make analysis writing more accessible by eliminating coding complexities, thereby allowing analysts with differing levels of computing skill to focus on the analysis design. It would allow an easier communication of the analysis content, thus making analysis validation and review easier. It would help to compare analyses and find overlaps, informing the analysis combination process.  Consequently, it would greatly facilitate interpretation of the experimental results within and outside the collaborations.  Moreover, the ability to describe an analysis in a framework-independent manner would simplify analysis preservation beyond the lifetime of both the analysis frameworks and the experiments.

All these potential benefits have motivated several efforts in the community for designing languages and tools to interpret them and run them on data.  A first dedicated workshop at Fermilab in 2019 brought together experimentalists, phenomenologists and computing experts to overview the existing efforts, discuss technical design details and draw a road map~\cite{LPCWS}.  In this note, we briefly review the current status of these existing languages and tools.

{\bf ADL and CutLang:} ADL (Analysis Description Language) is a domain-specific and declarative language derived from the early prototype language {\tt LHADA} (Les Houches Analysis Description Accord), initially designed by a group of experimentalists and phenomenologists~\cite{Brooijmans:2016vro, Brooijmans:2018xbu, Brooijmans:2020yij}. It consists of a plain, easy-to-read text file describing the analysis with syntax rules that include standard mathematical and logical operations and 4-vector algebra.  In the ADL file, object, variable, event selection definitions are clearly separated into blocks with a keyword value structure, where keywords specify analysis concepts and operations. Syntax includes mathematical and logical operations, comparison and optimization operators, reducers, 4-vector algebra and HEPspecific functions (e.g. $d\phi$, $dR$).  The ADL file is accompanied by a library of self-contained functions encapsulating variables variables with complex algorithms non-trivial to express with the ADL syntax (e.g. MT2, aplanarity) or non-analytic variables (e.g. efficiency tables, machine learning discriminators).  A repository of ATLAS and CMS analyses implemented in the ADL format is available in~\cite{ADLanalyses}.

Tools have been developed to interpret ADL and run the analysis it describes on events.  {\tt adlt2nm} is a python transpiler converting ADL to {\tt c++} code executed within the {\tt TNM} ({\tt TheNtupleMaker}) generic ntupling and analysis framework. It only depends on ROOT, and can work with any simple ntuple format, where input event formats are automatically incorporated into the {\tt c++} code~\cite{Brooijmans:2018xbu, adl2tnm}.  It assumes that a standard extensible type is available to model all analysis objects, and uses adapters to translate input to standard types.  {\tt CutLang} is a runtime interpreter written in {\tt c++} and based on ROOT~\cite{Sekmen:2018ehb, Unel:2019reo, cutlang}.  It parses ADL by {\tt Lex \& Yacc}, relying on automatically generated dictionaries and grammar.  {\tt CutLang} can process multiple input formats like {\tt Delphes}~\cite{Delphes}, CMS NanoAOD~\cite{nanoaod}, ATLAS and CMS open data~\cite{opendata} ntuples and FCC, while more can be easily added.  {\tt CutLang} additionally allows defining histograms within the ADL file. Both tools provide output in ROOT files, which include analysis algorithms, cutflows and counts and for {\tt CutLang}, user-defined histograms.  A parser from LHADA, the earlier prototype of ADL, to Rivet, called {\tt lhada2rivet} is also being developed~\cite{Brooijmans:2018xbu, lhada2rivet}.

{\bf F.A.S.T. analysis tools based on YAML:} {\tt F.A.S.T.} is a system that performs analysis using a YAML-based, naturally declarative DSL called {\tt YADL}~\cite{FAST}.  Analyses are run on data tables created from ROOT trees by YAML-based analysis configuration files handled and executed by python.  Data processing is described in stages, defined as python-importable classes.  {\tt F.A.S.T.} currently uses {\tt AlphaTwirl} as back-end, and can be developed to work with {\tt SPark} or {\tt RDataFrame}.  Variable definitions combine {\tt uproot} with {\tt numexpr} and work with jagged arrays.  Event selections are specified as nested dictionaries.  Histogram filling is also available.  F.A.S.T outputs a cutflow table with raw and weighted yields, provided as {\tt pandas dataframes}. It is used in 2 CMS analyses, and others in DUNE, FCC and LUX-ZEPLIN.

{\bf NAIL (Natural Analysis Implementation Language): } {\tt NAIL} is an embedded DSL written in python, that expresses event processing operations in a declerative form~\cite{NAIL}.  Its syntax resembles that of ROOT's {\tt RDataFrame}, in addition to which, it can view objects as a whole, select object subcollections, define new object properties, define operations between collections. It takes CMS {\tt NanoAOD}~\cite{nanoaod} as input data format. {\tt NAIL} generates a ROOT {\tt RDataFrame}-based code, either as a c++ program to compile and run, or a c++ library loadable with ROOT, even in a python environment.  It autodetects inputs while writing c++ code and allows to express new variable definitions with c++ code snippets.  {\tt NAIL} is currently being used in the CMS $H \rightarrow \mu \mu$ analysis~\cite{Sirunyan:2020two}.  Its development has started recently, and its ultimate scope will include defining event processing in a declarative way, building signal and background models from a set of histograms and defining statistical interpretation.

{\bf hep\_tables and dataframe\_expressions: } These are two packages that work together to allow easy columnar-like access to hierarchical data when the data processing back-end is in an analysis facility and the user is using either Jupyter notebooks or python source files to drive the analysis~\cite{heptables}.  Experimental data is processed by {\tt ServiceX}, a distributed cloud application that extracts columns of data quickly from it.  Analysis on data involves {\tt qastle}, a lisp-like language used to specify columns and simplified cuts.  The analysis is performed by {\tt heptables}, that converts a computational graph into commands to {\tt ServiceX} and {\tt awkward} and {\tt coffea} to do the user's bidding. User manipulations on data is allowed by {\tt dataframe\_expressions}.  The system performs single object and event selection cuts using {\tt numpy}-like slicing/masks.  It uses map and lambda functions for multiple object combinations and also have basic plotting functionality.  

{\bf PartiQL, AwkwardQL: } {\tt PartiQL} is a toy language demonstrating features that would be a radical departure from general purpose languages, addressing problems specific to particle physics~\cite{partiql}. It is intended to show how SQL's data model can improve data-handling in particle physics if applied to individual events, but not to collections of events. It has a block structure for named, nested cuts and systematic variations. Histograms are treated as side-effects within the block structure.  Identities are defined by surrogate keys.  Manipulation is done in terms of set operations, but not functions, and syntactic macros are used to avoid repetitive typing.  {\tt AwkwardQL} is an extension of {\tt PartiQL}, which is currently being designed to perform set operations on data expressed as awkward arrays~\cite{awkwardql}.  It will be focusing on more concrete physics cases.

Analysis description languages would greatly help to maximize the physics output of the LHC.  Several alternative approaches are being studied to design such languages and develop tools to interpret and run them on data.  These studies have proved the feasibility of the analysis description language approach, however there are still many intriguing problems to solve.  Developments in this are ongoing based on valuable input from implementations of real life analyses from the LHC experiments.  Discussions will continue in platforms like HSF, IRIS-HEP, Les Houches and LHC (Re)interpretation Forum, for getting feedback and ideas from a wide community, in order to progress towards the ultimate aim of a common analysis description language for the LHC that combines the best ideas from the different approaches.

\end{document}